\documentclass[aps,pra,prabib,twocolumn,showpacs,nofootinbib]{revtex4}
\usepackage{graphicx} \usepackage{amsmath} \usepackage{amssymb}
\usepackage{amsfonts} \usepackage{bm}
\usepackage{hyperref}

\begin{document}

\newcommand{\be}{\begin{equation}} \newcommand{\ee}{\end{equation}}
\newcommand{\bea}{\begin{eqnarray}}\newcommand{\eea}{\end{eqnarray}}

\title{Inverse square problem and  $so(2,1)$ symmetry in noncommutative  space}

\author{\href{http://pulakgiri.googlepages.com/home}{Pulak Ranjan Giri}} 
\email{pulakranjan.giri@saha.ac.in}

\affiliation{Theory Division, Saha Institute of Nuclear Physics,
1/AF Bidhannagar, Calcutta 700064, India}

\begin{abstract}
We study the quantum mechanics of a system with inverse square
potential in noncommutative space. Both the coordinates and
momentums are considered to be noncommutative, which breaks the
original $so(2,1)$ symmetry. The energy levels and eigenfunctions
are obtained. The generators of the $so(2,1)$ algebra are also
studied in noncommutative phase space and the commutators are
calculated, which shows that the $so(2,1)$ algebra obtained in
noncommutative space is not closed. However the commutative limit
$\Theta,\overline{\Theta}\to 0$ for the algebra smoothly goes to the
standard $so(2,1)$ algebra.
\end{abstract}


\date{\today}

\maketitle
%\section{Introduction}
 Noncommutative quantum mechanics has received lot of attention in
recent years
\cite{gamboa,gamboa1,gam,hov1,hov2,hov3,hov4,hov5,nair,jonke,rban,rban1,rban2,
chai1,bellucci3,bellucci4,bellucci41,bellucci42,barbosa,daval,muthu,gieres}. It was Snyder who
introduced the concept of noncommutative spacetime in his work
\cite{snyder,snyder1} in 1947, where electrodynamics is treated in
noncommutative spacetime. It is well known that the configuration
space of a  quantum mechanical system, confined to lowest Landau
level, behaves as phase space and thus becomes noncommutative. In
quantum mechanics noncommutativity is a property of the phase space
and this lead to the Heisenberg uncertainty relation $\Delta x\Delta
p_x\geq \hbar/2$. Coordinate space noncommutativity is much explored
and actively being studied nowadays in diverse fields
\cite{michael,calmet}.  The list includes  quantum mechanics
\cite{dayi,jella2,dadic,dayi1,jella1,kara,agni,bellucci2,chai,ani},
quantum field theory
\cite{sachin1,sachin2,bal1,rba,car1,falomir1,madore,lopez1,
calmet1,panero1,panero2,saha}, 
string
theory \cite{seiberg,chu1,chu2,doug} etc.

However momentum space noncommutativity  in
addition to coordinate space noncommutativity seems to be not much
explored although discussed recently in some literatures 
\cite{bertolami,jian}. It is however known that in background magnetic field,
$\boldsymbol{B}= (B^1, B^2, B^3)$, the different components of a
generalized momenta do not commute,
$\left[P^i,P^j\right]=i\epsilon^{ijk}B^k$. Note that the momentum
noncommutativity due to background magnetic field is governed by the electric
charge of the corresponding quantum particle. On the other hand we introduce
momentum noncommutativity which is independent of electric charge. 
In this article we thus
discuss a system with inverse square potential
\cite{giri,giri1,giri2,giri3,giri4,camblong1,camblong2} in
noncommutative momentum space in addition to coordinate space 
noncommutativity. Inverse square potential is very
important because of its large range of applicability. Starting from
electron scattering in polar molecules \cite{giri3} to scalar field dynamics in
near horizon space of many black hole spacetimes can be described by
Schr\"{o}dinger eigenvalue equation with inverse square potential.
It is interesting from the theoretical point of view also. It
belongs to a class of interactions which have conformal symmetry. Due
to this symmetry the system does not have any scale and thus it does
not possesses any finite ground state. Usually this system can be made
physical from the bound state point of view by a suitable
self-adjoint extensions
\cite{reed,kumar1,kumar2,kumar3,kumar4,kumar5,kumar6,kumar7,feher}.
Then the system possesses a finite ground state, which breaks the scale
symmetry. Symmetry breaking in the process of quantization is called
anomaly and inverse square interaction is thus a simple realization
of scaling anomaly. Renormalization \cite{rajeev} is  another
technique by which this inverse square potential can be treated.

In a recent paper \cite{pulak} we studied the inverse square system in only
noncommutative coordinate space. It was found that the original
$so(2,1)$ symmetry is broken by the scale $\Theta$ of the
noncommutative space. The  scale symmetry is explicitly broken by a
potential of the form $\sim 2\Theta\alpha r^{-4}L_z$ (to lowest order
in noncommutativity), which is generated due to the noncommutativity of
the coordinate space. This system then possesses a bound state at
threshold \cite{mako,jamil,hojman}, i.e., $E=0$. The system was also
studied for large noncommutative parameter $\Theta$ and the bound
state solutions are obtained, which however does not have the
commutative limit.

The article is organized in the following fashion: we first briefly
review the quantum mechanical system with inverse square potential
in a noncommutative plane, which will set the platform for our next 
discussion. 
Then we come to the discussion of our
present article and study  the same system in noncommutative
momentum space and study its solution and symmetry algebra. We also discuss
the system where both the coordinate-coordinate and momentum-momentum
noncommutativity  is taken into account at a time. Finally
we conclude.

Quantum mechanical system defined by the Hamiltonian $H$ with the
potential $V= \alpha r^{-2}$ is considered on noncommutative plane.
In noncommutative plane, the standard algebra on phase space gets
modified (we use the unit $\hbar=1$)
\begin{eqnarray}
\nonumber\left[\overline{x_i},\overline{x_j}\right]&=&
2i\epsilon_{ij}\Theta,~
\left[\overline{p_i},\overline{p_j}\right]=0,~
\left[\overline{x_i},\overline{p_j}\right]=i\delta_{ij}\,,\\
\epsilon_{12}&=&-\epsilon_{21}=1,
\epsilon_{11}=\epsilon_{22}=0\,.\label{al1}
\end{eqnarray}
However, in the commutative limit $\Theta\to 0$, the algebra
(\ref{al1}) reduces to the well known algebra on phase space
\begin{eqnarray}
\left[x_i,x_j\right]= 0,~\left[p_i,p_j\right]= 0,~
\left[x_i,p_j\right]=i\delta_{ij}\,.\label{al2}
\end{eqnarray}
One possible realization for the noncommutative phase space
coordinates
($\overline{x_i},\overline{p_i}, i=1,2$) in
terms of standard coordinates ($x_i, p_i, i=1,2$) is the following
\begin{eqnarray}
\nonumber \overline{x_1}&=&x_1 -\Theta p_2,~~ \overline{x_2}=x_2 +\Theta p_1\,,\\
\overline{p_1}&=&p_1,~~~~~ \overline{p_2}=p_2\,. \label{rep1}
\end{eqnarray}
It can be easily checked that the representation (\ref{rep1}) is
consistent with the algebra (\ref{al1}) and (\ref{al2}). Because of
the scale invariant potential $V$, the Hamiltonian $H$ possesses
$so(2,1)$ symmetry
\begin{eqnarray}
\left[D,H\right]= -iH,~~ \left[D,K\right]= iK,~~ \left[H,K\right]=
2iD\,,\label{algebra1}
\end{eqnarray}
in commutative plane ($x_i, i=1,2$). Here $D$ is the dilatation and $K$ is the
conformal operator. But when we consider the system
in noncommutative plane ($\overline{x_i}, i=1,2$), the scenario
changes completely. The $so(2,1)$ algebra up to first order in
$\Theta$ is modified like \cite{pulak}
\begin{eqnarray}
\nonumber \left[D_\Theta,H_\Theta\right]&=&-i H_\Theta
+\Theta\Delta_1\,, \left[D_\Theta,K_\Theta\right]=i K_\Theta +\Theta\Delta_2\,,\\
\left[K_\Theta, H_\Theta\right]&=&-2i D_\Theta+ \Theta \Delta_3
\,,\label{cm3}
\end{eqnarray}
where $\Delta_1= -2i\alpha r^{-4}L_z$, $\Delta_2= \left(2i\alpha
t^2r^{-4} +i/2\right)L_z$, $\Delta_3=-4i\alpha tr^{-4}L_z$. The
generators up to first order in $\Theta$ can be written as
\cite{pulak}
\begin{eqnarray}
H_\Theta= H + \Theta \Delta H, D_\Theta = D + \Theta \Delta D,
K_\Theta = K + \Theta\Delta K \,,\label{HDK3}
\end{eqnarray}
where $\Delta H=2\alpha r^{-4}L_z$, $\Delta D= 2\alpha r^{-4}L_zt$
and $\Delta K= (2\alpha t^2r^{-4}-1/2)L_z$, The above algebra
(\ref{cm3}), which is not closed, reduces to the well known
$so(2,1)$ algebra (\ref{algebra1}) in the limit $\Theta\to 0$.

We now come to the discussion of the system $H$ in noncommutative
momentum space. Since this time the momentums are noncommutative but
coordinates are commutative the algebra for the phase space becomes
\begin{eqnarray}
\left[\overline{x_i},\overline{x_j}\right]= 0,~
\left[\overline{p_i},\overline{p_j}\right]=2i\epsilon_{ij}\overline{\Theta},~
\left[\overline{x_i},\overline{p_j}\right]=i\delta_{ij}\,.\label{al3}
\end{eqnarray}
Here the strength of noncommutativity in momentum space is given by
the parameter $\overline{\Theta}$. The commutative limit
$\overline{\Theta}\to 0$ of the algebra (\ref{al3}) goes to the
standard  result (\ref{al2}). We choose one possible realization
for the noncommutative phase space coordinates
($\overline{x_i},\overline{p_i}, i=1,2$) in terms of standard
coordinates ($x_i, p_i, i=1,2$) as
\begin{eqnarray}
\nonumber \overline{x_1}&=&x_1 ,~~ \overline{x_2}=x_2 \,,\\
\overline{p_1}&=&p_1 +\overline{\Theta} x_2,~~~~ \overline{p_2}=p_2
-\overline{\Theta} x_1\,. \label{rep2}
\end{eqnarray}
Taking the representation (\ref{rep2}) the Hamiltonian
$H={\boldsymbol{p}}^2 +\alpha {\boldsymbol{r}}^{-2}$ can be written
in noncommutative momentum space as
\begin{eqnarray}
\nonumber H_{\overline{\Theta}}&=&{\boldsymbol{\overline{p}}}^2
+\alpha
{\boldsymbol{\overline{r}}}^{-2}\\
 &=&{\boldsymbol{p}}^2
+\alpha {\boldsymbol{r}}^{-2}
+\overline{\Theta}^2\boldsymbol{r}^2-2\overline{\Theta}
L_z\,.\label{pot1}
\end{eqnarray}
Once we write the Hamiltonian $H_{\overline{\Theta}}$ in terms of
the standard coordinates, we can solve the eigenvalue problem
\begin{eqnarray}
H_{\overline{\Theta}}\psi= E_{\overline{\Theta}}\psi\,.
\label{evalue}
\end{eqnarray}
This eigenvalue problem (\ref{evalue}) can be solved exactly  for
different ranges of the coupling constant $\alpha$, following a
method of Ref. \cite{feher}. The problem can be analyzed in 
different coupling constant ranges $\alpha\geq 1-m^2$, $-m^2 <
\alpha < 1-m^2$,  $\alpha <-m^2$ and $\alpha= -m^2$. It is well
known that for $\alpha\geq 1-m^2$, the Hamiltonian
$H_{\overline{\Theta}}$ is essentially self-adjoint and has unique
self-adjoint extensions. The bound state solutions and eigenvalues
are
\begin{eqnarray}
E_{\overline{\Theta}}^{n,m}&=& 2\overline{\Theta}\left[2n-m
+\sqrt{m^2+\alpha} +1\right]
 \label{sola}\,,\\
R_{\overline{\Theta}}^{n,m}(r)&=& r^{\sqrt{\alpha
+m^2}}e^{-\frac{1}{2}\overline{\Theta} r^2}L_n^{\sqrt{\alpha
+m^2}}(\overline{\Theta} r^2)\,, \label{sol}
\end{eqnarray}
where $L_n^{\sqrt{\alpha +m^2}}$ is Laguerre polynomial, $n=
0,1,2,... $ and $m= 0,\pm 1,\pm 2,....$. The emergence of these bound
states with eigenvalues $E_{\overline{\Theta}}^{n,m}$ are the
consequences of  the breaking of $so(2,1)$ symmetry due to the scale
$\overline{\Theta}$. For $-m^2 < \alpha < 1-m^2$, there is a one
parameter family of self-adjoint extensions. The exact solution for
a fixed self-adjoint extension parameter can be evaluated numerically. 
But for some
special values of the extension parameter analytical results can be
obtained also \cite{feher}. Among the two analytical results one
coincides with the results (\ref{sola}) and  (\ref{sol}) and the
other is given by
\begin{eqnarray}
E_{\overline{\Theta}}^{n,m}&=& 2\overline{\Theta}\left[2n-m
-\sqrt{m^2+\alpha} +1\right]
 \label{sola1}\,,\\
R_{\overline{\Theta}}^{n,m}(r)&=& r^{-\sqrt{\alpha
+m^2}}e^{-\frac{1}{2}\overline{\Theta} r^2}L_n^{-\sqrt{\alpha
+m^2}}(\overline{\Theta} r^2)\,. \label{sol5}
\end{eqnarray}
It should be noted that despite the constraint  $0 < \alpha + m^2 <
1$, all energy levels may not be positive here unlike the case of
Ref. \cite{feher} where all energy levels for the same constraint
take positive values. Depending upon the values of $m$, the energy
levels (\ref{sola1}) associated with high angular momentum may also
take negative values. For $\alpha <-m^2$ the energy levels are
unbounded from below and for $\alpha=- m^2$ the discussion for the
eigenvalue and eigenfunction can also be discussed in line with
Refs. \cite{giri,giri1,giri2}.

It would be informative to study the $so(2,1)$ algebra in
noncommutative momentum space. To the first order of the
noncommutativity $\overline{\Theta}$ the algebra becomes
\begin{eqnarray}
\left[D_{\overline{\Theta}},H_{\overline{\Theta}}\right]=-i H,
\left[D_{\overline{\Theta}},K_{\overline{\Theta}}\right]=i K,
\left[K_{\overline{\Theta}}, H_{\overline{\Theta}}\right]=-2i D
,\label{cm3m}
\end{eqnarray}
which is easily seen to be reduced to the standard $so(2,1)$ algebra
(\ref{algebra1}) in the limit $\overline{\Theta}\to 0$.

We now consider the phase space where both the position space and
momentum space are taken to be noncommutative. The noncommutative
algebra on the phase space,  obtained from (\ref{al1}) and
(\ref{al3}), can be written as
\begin{eqnarray}
\left[\overline{x_i},\overline{x_j}\right] =
2i\epsilon_{ij}\Theta,~
\left[\overline{p_i},\overline{p_j}\right]=2i\epsilon_{ij}\overline{\Theta},~
\left[\overline{x_i},\overline{p_j}\right]=i\delta_{ij}\,.\label{al4}
\end{eqnarray}
It is possible to get the noncommutative phase space coordinates in
terms of standard coordinates. From (\ref{rep1}) and (\ref{rep2}) we
get the following representation
\begin{eqnarray}
\nonumber \overline{x_1}&=&x_1 -\Theta p_2,~~ \overline{x_2}=x_2 +\Theta p_1\,,\\
\overline{p_1}&=&p_1 +\overline{\Theta} x_2,~~\overline{p_2}=p_2
-\overline{\Theta} x_1\,,
 \label{rep3}
\end{eqnarray}
which is consistent with the algebra (\ref{al4}) up to the first
order in noncommutative parameters $\Theta$ and $\overline{\Theta}$.
Note that the representation (\ref{rep3}) satisfies the first two
commutators of the  algebra (\ref{al4}) to all orders in
noncommutative parameters $\Theta$ and $\overline{\Theta}$. But the
last commutator is only satisfied for the first order in
noncommutativity. One can also consider the all higher orders of the
noncommutative parameters, which will give \cite{bertolami} (we
explicitly keep $\hbar$ here)
\begin{eqnarray}
\left[\overline{x_i},\overline{p_j}\right]=i\delta_{ij}
\left[1+\Theta\overline{\Theta}\right]\hbar\,.\label{al5}
\end{eqnarray}
Note that if we ignore terms higher than  the first order  in
noncommutative parameters then the second term within the bracket in
right side of the equation (\ref{al5}) should be ignored and the
resulting equation is the standard expression. The fact that the
Plank constant gets modified as \cite{bertolami,branko}
\begin{eqnarray}
\overline{\hbar}=
\left[1+\Theta\overline{\Theta}\right]\hbar\,,\label{al6}
\end{eqnarray}
is a exclusive feature of the noncommutative phase space, which was
absent both in noncommutative coordinate space and noncommutative
momentum space. In Ref. \cite{bertolami} this effective Plank
constant has been exploited to calculate a bound on the
noncommutative parameters.

We now concentrate on the system in noncommutative phase space. The
Hamiltonian in this situation becomes
\begin{eqnarray}
\nonumber H_{\Theta,\overline{\Theta}}={\boldsymbol{\overline{p}}}^2
+\alpha
{\boldsymbol{\overline{r}}}^{-2} \hspace{3cm}\\
 ={\boldsymbol{p}}^2 +\overline{\Theta}^2\boldsymbol{r}^2-2\overline{\Theta}
L_z + \hspace{1cm}\nonumber \\
\frac{\alpha}{\left(\Theta^2\boldsymbol{p}^2 +
\boldsymbol{r}^2-2\Theta L_z\right)} \,.\label{pot2}
\end{eqnarray}
This Hamiltonian can be solved for large $\Theta$  using an algebraic method
followed in Ref. \cite{pulak}. For the moment we consider the
denominator of the last term of the Hamiltonian
$H_{\Theta,\overline{\Theta}}$,
\begin{eqnarray}
\overline{H_\Theta}= \Theta^2{\boldsymbol{p}^2}+r^2-2\Theta L_z\,,
\label{largetheta}
\end{eqnarray}
and write it in terms of the Schwinger representation. The
annihilation operators \cite{agni}
\begin{eqnarray}
\nonumber \overline{a_+}=(x_1-ix_2)+\Theta(ip_1+p_2)\,,\\
\overline{a_-}=(ix_1-x_2)-\Theta(p_1+ip_2)\,,
\label{representation1}
\end{eqnarray}
and its corresponding creation operators satisfy the  commutation
relation
\begin{eqnarray}
\left[\overline{a_+},\overline{a_+}^\dagger\right]=
\left[\overline{a_-},\overline{a_-}^\dagger\right]=4\Theta\,.
\label{representation2}
\end{eqnarray}
Rest of the  commutators  are  zero. The number operators can now be
constructed as
\begin{eqnarray}
\overline{n_+}= \overline{a_+}^\dagger
\overline{a_+}\,,~~~\overline{n_-}= \overline{a_-}^\dagger
\overline{a_-}\,,\label{number}
\end{eqnarray}
which satisfy the eigenvalue equation
\begin{eqnarray}
\nonumber \overline{n_+}|n_+,n_-\rangle= n_+|n_+,n_-\rangle\,,
n_+=0,4\Theta,8\Theta,12\Theta,...\\\overline{n_-}|n_+,n_-\rangle=
n_-|n_+,n_-\rangle\,,
n_-=0,4\Theta,8\Theta,12\Theta,...\label{number1}
\end{eqnarray}
The Hamiltonian  $\overline{H_\Theta}$ can now be written in terms
of the number operators as,
\begin{eqnarray}
\overline{H_\Theta}= \overline{n_-} + 2\Theta\,, \label{nham2}
\end{eqnarray}
which satisfy the  equation
\begin{eqnarray}
\overline{H_\Theta}|n_+,n_-\rangle =
\overline{E_\Theta}|n_+,n_-\rangle\\ \overline{E_\Theta}=
n_-+2\Theta\,. \hspace{1cm}\label{nham3}
\end{eqnarray}
Now the eigenvalue of the Hamiltonian $H_{\Theta,\overline{\Theta}}$
in $|n_+,n_-\rangle$ basis becomes
\begin{eqnarray}
\nonumber E_{\Theta,\overline{\Theta}}=\hspace{6cm}\\
\nonumber\langle n_+,n_-|\boldsymbol{p}^2
+{\overline{\Theta}}^2r^2-2\overline{\Theta}L_z|n_+,n_-\rangle
+\frac{\alpha}{n_-+2\Theta}\\\nonumber =
\frac{1}{4\Theta^2}\left[n_++n_-+4\Theta\right]+\frac{{\overline{\Theta}}^2}{4}
\left[n_++n_-+4\Theta\right]\\-\frac{\overline{\Theta}}{2\Theta}\left[n_+-n_-\right]
+\frac{\alpha}{n_-+2\Theta}\,. \hspace{1cm}\label{largen}
\end{eqnarray}
One should note that although the limit $\Theta\to 0$ can not be
taken in (\ref{largen}) directly but the limit $\overline{\Theta}\to
0$ can be taken smoothly and the limit goes to the result,  Eq.
(24), of Ref. \cite{pulak}. It is obvious that the existence of the
eigenvalue (\ref{largen}) is a consequence of the breaking of the
$so(2,1)$ symmetry. It is therefore interesting to get an explicit
commutator relations of the three generators of the $so(2,1)$
algebra in noncommutative phase space, which is up to first order in
noncommutative parameters $\Theta$ and $\overline{\Theta}$,
\begin{eqnarray}
\left[D_{\Theta,\overline{\Theta}},H_{\Theta,\overline{\Theta}}\right]&=&-i
H_\Theta
+\Theta\Delta_1\,,\\
 \left[D_{\Theta,\overline{\Theta}},K_{\Theta,\overline{\Theta}}
 \right]&=&i K_\Theta +\Theta\Delta_2\,,\\
\left[K_{\Theta,\overline{\Theta}},
H_{\Theta,\overline{\Theta}}\right]&=&-2i D_\Theta+ \Theta \Delta_3
\,.\label{cm4}
\end{eqnarray}
Note that the right hand side depends only on the coordinate space
noncommutative parameter $\Theta$, where we  consider only first order
terms. Note also that the commutative limit
$\Theta,\overline{\Theta}\to 0$ goes to the standard $so(2,1)$
algebra.

In conclusion, the inverse square potential is studied in
noncommutative space. Both the coordinates and momentums are
considered to be noncommutative. The bound state solutions are
obtained, which is a consequences of the scale symmetry breaking by
the noncommutative parameters $\Theta$ and $\overline{\Theta}$. The
three generators of the $so(2,1)$ algebra are also studied and the
commutators are constructed to first order in noncommutativity,
which shows that the algebra is not closed. However the algebra
reduces in the commutative limit $\Theta,\overline{\Theta}\to 0$ to standard
$so(2,1)$ algebra.

\end{document}